\documentclass[twocolumn]{revtex4}
\usepackage{amssymb}
\usepackage{latexsym}
\usepackage{epsfig}

\usepackage{amsmath}
\begin{document}
\title{New Tsallis agegraphic Dark Energy in Chern-Simons modified gravity }
\author{M. Abdollahi Zadeh \footnote{
mazkph@gmail.com}}
\address{ Not affiliated with any institution, Kazerun, Iran.}

\begin{abstract}
In this paper, we explore various cosmological parameters and cosmological planes ($\omega_D-{\omega_D}^{\prime}$, statefinder) for  new Tsallis agegraphic dark energy in the framework of Chern-Simons modified gravity. We consider this scenario in a flat FRW universe for both non-interacting and interacting cases, between dark energy and dark matter. It is observed that the equation of state parameter gives quintessence-like nature of the universe in all of the cases. Also, the squared of  the sound speed shows unstable behaviour for this scenario. The $\omega_D-{\omega_D}^{\prime}$ plane represents the freezing region in most of the cases.

\end{abstract}
\maketitle
\section{Introduction}
The surprising discovery of the accelerated expansion of the Universe is one of the exciting progress in cosmology in the last few years \cite{Riess,Komatsu,Ichiki}. However, a search for the exact nature of the phenomenon driving this acceleration is still under way. In order to explain this behavior, two main approches are considered: introducing the concept of dark energy (DE) as a new mysterious cosmological component or modifying the gravitational part of the Einstein equations\cite{Randall,Randall1,Dvali}.\\The Chern-Simons modified gravity theory is one of modified theories of gravity that has been developed for explaining the accelerated expansion, which is motivated from string theory and loop quantum gravity. This modified gravity theory is an extension to GR, introduced by Jackiw and Pi \cite{Jackiw}, aimed to solve the long standing problem of cosmic baryon asymmetry by introducing a parity-violating CS term to the Einstein-Hilbert action. Here, the gravitational field is coupled with a scalar field using a parity-violating Chern-Simons term. Many authors have investigated cosmological implications of dark energy models in dynamical Chern-Simons modified gravity. For example, cosmological solution of Chern-Simons modified gravity with Ricci-Gauss-Bonnet holographic dark energy model has been discussed in \cite{star1}. The pilgrim
dark energy model (PDE) with Hubble and event horizons and Ricci dark energy model in the framework
of dynamical Chern-Simons modified gravity have also been checked in \cite{star2, star3}, respectively. The new holographic dark energy model  and interacting new holographic dark energy in dynamical Chern-Simons gravity have been studied in \cite{star4, star8} respectively, as well as, cosmological evolution of modified QCD ghost dark energy
in dynamical Chern-Simons gravity is examined \cite{star5}. Holographic dark energy models and higher order generalizations
in dynamical Chern-Simons modified gravity have been studied in \cite{star6}. A study of the generalized
entropy based holographic dark energy
models in dynamical Chern-Simons modified
gravity has been done in \cite{star7}. For more review on cosmic acceleration in the dynamical Chern-Simons modified gravity can see \cite{star9, star10, star11}.
 \\ Among all theoritical models, $\Lambda CDM$ model, which consists of a mixture of the cosmological constant $\Lambda$ and the cold dark matter (CDM) has most compatible with observational data. However, this model suffers from some problems such as the fine-tuning and the coincidence problems \cite{Ros25}. In order to solve those two problems, many dynamical dark energy models have been proposed namely, agegraphic dark energy\cite{Cai}, which is based on the karolyhazy uncertainly relation\cite{K´arolyh´azy}
\begin{eqnarray}
\delta{t}=\beta t_{p}^{2/3}t^{1/3}
\end{eqnarray}
where $\beta$ is a dimensionless constant of order unity and
$t_{p}$ denotes the reduced Plank time. Since, this model suffers from the contradiction to describe the matter-dominated universe in the fast past, the new agegraphic dark energy model with the conformal time $\eta$ is proposed by Wei and Cai \cite{Wei}.\\ As Gibbs has pointed out, in systems in which the partition function diverses the Boltzman-Gibbs theory cannot be applied and the Boltzman-Gibbs additive entropy must be generalized to the non-additive entropy, known as Tsallis entropy\cite{Tsallis,Tsallis1,Tsallis2}. On the other hand, Tsallis and Cirto\cite{Cirto}, stated that, the entropy associated with the black hole is written as $S_{\delta}=\gamma A^{\delta}$, where $\gamma$ is an unknown constant and $\delta$ denotes the non-extensive parameter. As we know, for the gravitational
systems like, cosmic lies within this class, it may be better that  have described by using the
generalized entropies namely Tsallis entropy. In this regard, applying Tsallis generalized entropy in holographic hypothesis led to Tsallis holographic dark energy (THDE). A number of papers on dark energy models based on Tsallis entropy are available in Refs \cite{Tavayef2,Tavayef3,Abdollahi,Tavayef10,Abdollahi40,Tavayef4,Abdollahi41,Tavayef5,Tavayef6,Tavayef8,Tavayef9,Tavayef11,Tavayef12,Tavayef13,Tavayef15,Tavayef16,Tavayef17,Tavayef18,Tavayef19,Tavayef20,Tavayef21,Tavayef22,Tavayef23,Tavayef24}. Based on generalized Tsallis entropy formalism, the interacting and non-interacting DE models are explored with different IR cutoffs \cite{Tavayef2}. Thermal stability of THDE model with apparent horizon as IR cutoff has been studied in Ref \cite{Tavayef3}. The cosmological features of Tsallis agegraphic dark energy (TADE) models assuming the age of the Universe and the conformal time as the IR cutoffs have been investigated \cite{Abdollahi}. The effects of anisotropy on THDE  model, assuming sign-changeable interaction between DM and THDE are explored with different IR cutoffs \cite{Tavayef10}. The Tsallis agegraphic dark energy model by considering a sign- changeable interaction between TADE and DM has been studied \cite{Abdollahi40}. Authors \cite{Tavayef4}, have investigated the ThDE in cyclic, DGP and RSII braneworld models with the Hubble radius as the IR cutoff. The cosmological consequences of Tsallis, Renyi and Sharma-Mittal holographic and new agegraphic dark energy models in the context of D-dimensional fractal universe have been studied \cite{Abdollahi41}.

The cosmological consequences of  the generalized
entropy based  HDE
models in the dynamical
Chern-Simons framework, as a modified gravity theory, can be found in Ref \cite{Tavayef25}. 
 In this paper, we want to study the nature of the new agegraphic dark energy by assuming Tsallis entropy in framework Chern-Simons modified gravity. \\ We organized the paper as follows. In section II, we provide the basic cosmological scenario of Chern-Simons modified gravity. In section III, we describe new Tsallis  agegraphic dark energy, statefinder plane,  $\omega_D-{\omega_D}^{\prime}$ plane and stability analysis. In section IV, we study cosmological parameters and cosmological planes for both non-interacting and interacting senarios in NTADE at background CS modifieg gravity. In the last, we state our results.


\section{Chern-Simons modified gravity}
The dynamical Chern-Simons modified gravity can be described by the following action \cite{Alexander}
\begin{equation}\label{1}
S=\frac{1}{16 \pi G}\int_{\nu}d^{4}x\left[\sqrt{-g}R+\frac{l}{4} \theta {^\ast R}^{\rho\sigma\mu\nu}R_{\rho\sigma\mu\nu}+L1
\right]+S_{mat},
\end{equation}
\begin{equation}\nonumber
L1=-\frac{1}{2}g^{\mu\nu}\nabla_{\mu}\theta\nabla_{\nu}\theta +V(\theta)
\end{equation}
where $R$ represents the Ricci scalar, $\nabla_{\mu}$ is the covariant derivative, $\theta$ is the dynamical variable, $l$ shows a coupling constant, ${^\ast R}^{\rho\sigma\mu\nu}R_{\rho\sigma\mu\nu}$
is a topological invariant called the Pontryagin term, $V(\theta)$ is the potential term that is set to zero in the current work for simplicity and  $S_{mat}$ represents the action of matter.
Now, the variation of the action with respect to metric tensor $g_{\mu\nu}$ and the scalar field $\theta$, yields two field equations of CS modified gravity 
\begin{eqnarray}
G_{\mu\nu}+l C_{\mu\nu}&=&8 \pi G T_{\mu\nu},\nonumber \\
\label{2}g^{\mu\nu}\nabla_{\mu}\nabla_{\nu}\theta&=&-\frac{l}{64 \pi}{^\ast R}^{\rho\sigma\mu\nu}R_{\rho\sigma\mu\nu},
\end{eqnarray}
where, $G_{\mu\nu}$ appears as Einstein tensor and $C_{\mu\nu}$ is known as the Cotton tensor which is defined as 
\begin{equation}\label{3}
C_{\mu\nu}=-\frac{1}{2\sqrt{-g}}((\nabla_{\rho}\theta)\varepsilon^{\rho \beta \tau (\mu} \nabla_{\tau} R^{\nu)}_{\beta})
+(\nabla_{\sigma}\nabla_{\rho}\theta)^{\ast} {R}^{\rho(\mu\nu)\sigma}.
\end{equation}
Moreover, the energy-momentum tensor $T_{\mu\nu}$, is composed of two parts, the matter part $T^{m}_{\mu\nu}$ and the scalar field part ${\hat{T}^{\theta}}_{\mu\nu}$ as
\begin{eqnarray}
{\hat{T}^{\theta}}_{\mu\nu}&=&\nabla_{\mu}\theta \nabla_{\nu}\theta-\frac{1}{2}g_{\mu\nu}\nabla^
{\rho}\theta\nabla_{\rho}\theta,\nonumber \\
\label{4}T^{m}_{\mu\nu}&=&(\rho+p)u_{\mu}u_{\nu}+pg_{\mu\nu},
\end{eqnarray}
where $P$ and $\rho$ represent the pressure and
energy density respectively. Furthermore, $u$ is the four-vector velocity in co-moving coordinates of the spacetime. In the framework of Chern-Simons modified gravity, for flat FRW universe, the first Friedmann equation can be obtained by using Eqs.(\ref{2}) and (\ref{4}) as follows 
\begin{equation}\label{5}
H^{2}=\frac{1}{3}(\rho_{m}+\rho_{D})+\frac{1}{6}\dot{\theta}^{2}.
\end{equation}
Here, the dot denotes the derivative of scalar factor $a$ with respect to cosmic time and $m^{-1}_{pl}=8\pi G=1$. The FRW metric yields ${^\ast R}^{\rho\sigma\mu\nu}R_{\rho\sigma\mu\nu}=0$, hence the field equation (\ref{2}) is associated with the scalar field takes the following form
\begin{equation}\label{6}
g^{\mu\nu}\nabla_{\mu}\nabla_{\nu}\theta=g^{\mu\nu}[\partial_{\nu}\partial_{\mu}\theta]=0.
\end{equation}
With the assumption $\theta=\theta(t)$, the above equation leads to 
\begin{equation}\label{7}
\ddot{\theta}+3H\dot{\theta}=0,
\end{equation}
which implies that $\dot{\theta}=c a^{-3}$, $c$ is a constant of integration. In this way, we obtain the expression for Eq.(\ref{5}) as 
\begin{equation}\label{8}
H^{2}=\frac{1}{3}(\rho_{m}+\rho_{D})+\frac{1}{6}{c^{2}a^{-6}}.
\end{equation}
In the framework of Chern-Simons modified gravity, the continutiy equation becomes 
\begin{eqnarray}
\dot{\rho}+3H(\rho+P)=0,
\end{eqnarray}
where by taking the interaction between dark matter and dark energy into account the above equation may be written as 
\begin{eqnarray}
\dot{\rho}_{m}+3H\rho_{m}&=&Q, \label{9}
\\\label{10}\dot{\rho}_{d}+3H(\rho_{D}+p_{D})&=&-Q.
\end{eqnarray}
and $Q$ is the interaction term and due to the continuity equation, it must be a function of the product of energy density and a term with unity of time (such as Hubble parameter). With this idea, from different forms that have been proposed for interaction, we take the following form as 
\begin{eqnarray}\label{Q}
Q=3 H b^{2}\rho_{m},
\end{eqnarray}
with $b^2$ as an interaction parameter which transfers the energy between DM and DE. Using this expression in Eq. (\ref{9}), we get 
\begin{equation}\label{11}
 \rho_{m}= \rho_{m 0}a^{-3(1-b^{2})},
\end{equation}
where, $\rho_{m0}$ is an integration constant.
\section{New Tsallis  agegraphic Dark Energy}
Due to some problems in agegraphic dark energy model, which is based on the uncertainty relation of quantum mechanics \cite{Cai} and the time scale $t$ is chosen to be IR cutoff, Wei and Cai \cite{Wei} proposed a new model of agegraphic dark energy, while the time scale is chosen to be the conformal time $\eta$ instead of the age of the universe, which is defined by $dt=ad\eta$, where $t$ is the cosmic time.\\ On the other hand, Tsallis and Cirto \cite{Cirto}, stated that, the entropy associated with the black hole is written as $S_{\delta}=\gamma A^{\delta}$, where $\gamma$ is an unknown constant and $\delta$ denotes the non-extensive parameter. Recently, applying Tsallis generalized entropy in holographic hypothesis, a new dark energy density as 
\begin{eqnarray}\label{Trho}
\rho_D=BL^{2\delta-4},
\end{eqnarray}
was proposed, where $B$ is an unknown parameter and $L$ is the IR cutoff. Now, with assuming $\eta=L$, we get the energy density of NTADE as \cite{Abdollahi}.
\begin{eqnarray}\label{age}
\rho_D=B{\eta}^{2\delta-4}.
\end{eqnarray}
\subsection{statefinder parameter, $\omega_D-{\omega}^{\prime}_{D}$
plane and classical stability}
Since numerous dynamical dark energy models is being developed to clarify the accelerating expansion of the universe, we need the geometrical statefinder operators which are used to undrestand the differences between different dark energy models. These general expansions are defined as \cite{Sahni}
\begin{equation}\label{rr2}
r=2q^2+q-\frac{\dot q}{H},
\end{equation}
\begin{equation}\label{statefinder}
 s=\frac{r-1}{3(q-1/2)},
\end{equation}
Actually, we should point to the fact that in the $\{r,s\}$ plane, $s >0$ corresponds to a quintessence- like model of DE and $s<0$ corresponds to a
phantom-like model of DE, while the point  $\{r,s\}=\{1,0\}$ represents $\Lambda$CDM. Another way for analyzing the dynamical property of various DE models and distinguish these models is using the $\omega_D-{\omega}^{\prime}_{D}$
plane, in which ${\omega}^{\prime}_{D}>0$ and $\omega_D<0$
 present the thawing region, while ${\omega}^{\prime}_{D}<0$
and $\omega_D<0$  present the freezing region \cite{Caldwell}. In order to check the stability of the DE models, we should evaluate the sign of the squared sound speed. If the sign of ${v}^{2}_{s}$, which is given by 
\begin{equation}\label{vs}
v_{s}^{2}=\frac{dP_D}{d\rho_D}=\frac{\dot{P}_D}{\dot{\rho}_D}=\dfrac{\rho_{D}}{\dot{\rho}_{D}}
\dot{\omega}_{D}+\omega_{D},
\end{equation}
to be positive, the model is stable otherwise it is unstable\cite{Peebles}
\section{new Tsallis agegraphic dark energy in cs modified gravity}
\subsection{Non-interacting case}
With assuming non-interacting case ($Q=0$), we obtain the cosmological parameters, which help us for understanding behaviour of this model. At first, by inserting Eq.(\ref{age}) and its time derivative into Eq.(\ref{10}), we get 
\begin{eqnarray}\label{EoSna}
\omega_D=-1-\frac{2\delta-4}{3a\eta H},
\end{eqnarray}
where $\eta=(\frac{3 H^2 \Omega_D}{B})^{\frac{1}{2\delta-4}}$. 
The expression for ${\omega}^{\prime}_{D}$ can obtain by taking the derivative of the EoS parameter (\ref{EoSna}) with respect to $x=lna$
\begin{equation}\nonumber
{\omega}^{\prime}_{D}=\frac{-3 a^{3}H_0^2\Omega_{m0} \eta^{3-2\delta}((\delta-2)(\Omega_D))}{a^7 B H}
\end{equation}
\begin{equation}\label{primeEoS}
+\frac{(\delta-2)(-c^2 \eta+2a^5 H(1+a\eta H+(\delta-2)\Omega_D))}{3a^7 \eta^2 H^3}.
\end{equation}

We can find the rate of universe expansion with calculation the deceleration parameter $q$, which is defined as 
\begin{equation}\label{qage}
q\equiv-1-\frac{\dot{H}}{H^2},
\end{equation}
by differentiating from Eq.(\ref{8}) and using Eq.(\ref{9}) and equation 
\begin{equation}\label{age1}
\dot{\rho_D}=\frac{2\delta-4}{a\eta}\rho_D,
\end{equation}
as
\begin{equation}\label{qnage}
q=-\frac{1}{3}-\frac{3\Omega_D}{2}-\frac{(\delta-2)\Omega_D}{a\eta H}+\frac{c^2a^{-6}}{2H^2}.
\end{equation}
In addition, we can obtain a mathematical expression for the fractional DE density as 
\begin{equation}\label{nageOmega}
\dot{\Omega}_D=\frac{(2\delta-4)\Omega_D}{a\eta}+2\Omega_D H(1+q).
\end{equation}

\begin{figure}[htp]
\begin{center}
\includegraphics[width=8cm]{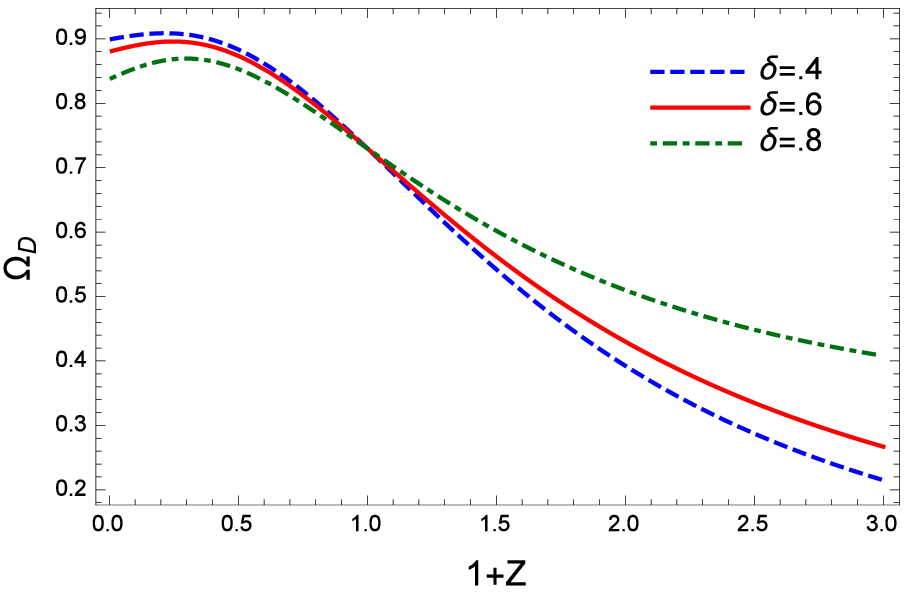}
\caption{The evolution of $\Omega_D$ versus redshift parameter $z$ for
 non-interacting NTADE in CS modified gravity. Here, we have taken
$\Omega_D(z=0)=0.73$, $H(z=0)=74$, $\Omega_{m0}=.23$, $c^2=.25$ and $B=2.4$.
}\label{Omega-z1}
\end{center}
\end{figure}

\begin{figure}[htp]
\begin{center}
\includegraphics[width=8cm]{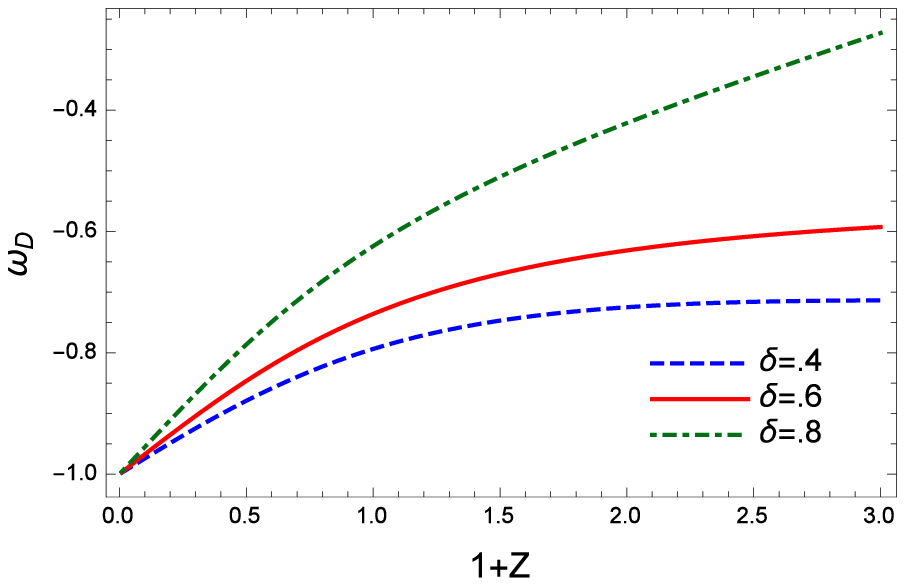}
\caption{The evolution of $\omega_D$ versus redshift parameter $z$ for
 non-interacting NTADE in CS modified gravity. Here, we have taken
$\Omega_D(z=0)=0.73$, $H(z=0)=74$, $\Omega_{m0}=.23$, $c^2=.25$ and $B=2.4$.}\label{w-z1}
\end{center}
\end{figure}

\begin{figure}[htp]
\begin{center}
\includegraphics[width=8cm]{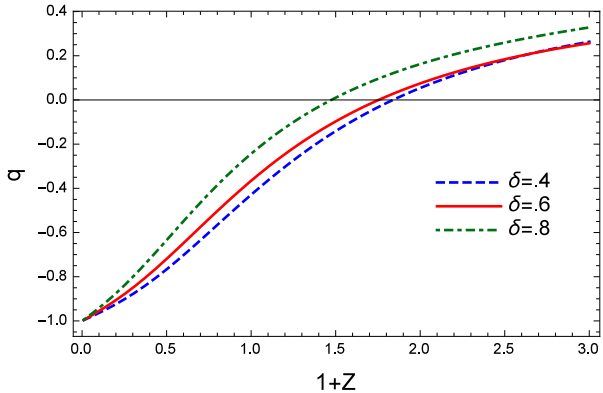}
\caption{The evolution of the deceleration parameter $q$ versus redshift parameter $z$ for
 non-interacting NTADE in CS modified gravity. Here, we have taken
$\Omega_D(z=0)=0.73$, $H(z=0)=74$, $\Omega_{m0}=.23$, $c^2=.25$ and $B=2.4$.}\label{q-z1}
\end{center}
\end{figure}

By taking the time derivative of Eq.(\ref{EoSna}) and using Eq.(\ref{age1}) and then replacing in relation (\ref{vs}), we obtain an expression for $v_{s}^{2}$ as 

\begin{equation}\nonumber
v_{s}^{2}=\frac{-9a^3 {H_0}^2 \Omega_{m0} \eta^{4-2\delta} \Omega_D}{6a^6 B}
\end{equation}
\begin{equation}\label{vs1}
+\frac{-c^2 \eta+2a^5 H(5-2\delta-2a\eta H+(\delta-2)\Omega_D)}{6 a^6 \eta H^2}.
\end{equation}

\begin{figure}[htp]
\begin{center}
\includegraphics[width=8cm]{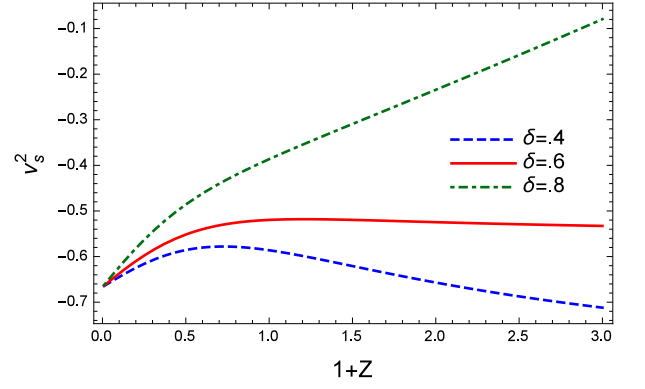}
\caption{The evolution of  the squared of sound speed $v_s^2 $ versus redshift parameter $z$ for
 non-interacting NTADE in CS modified gravity. Here, we have taken
$\Omega_D(z=0)=0.73$, $H(z=0)=74$, $\Omega_{m0}=.23$, $c^2=.25$ and $B=2.4$.}\label{v-z1}
\end{center}
\end{figure}

Using the time derivative of Eq. (\ref{qnage}), we can find the $\{r,s\}$ pair as the statefinder parameters.
\begin{equation}
r=M1+M2.
\end{equation}
\begin{equation}\nonumber
M1=\frac{9H{H_0}^2 \Omega_{m0}(\delta-2)\eta^5 \Omega_D^{2}-2B \eta^{2\delta}a^{2}(\delta-2)^{2}\Omega_D^2}{2BH^{2}a^{4}\eta^{2(1+\delta)}},
\end{equation}
\begin{equation}\nonumber
M2=\frac{B\eta^{2\delta}\left(aH\eta^2(3c^22a^6H^2)+m2\right)}{2B H^3 a^7 \eta^{2(1+\delta)}},
\end{equation}
\begin{equation}\nonumber
m2=+(\delta-2)(c^2\eta+2Ha^5(-5+2\delta+2a \eta H))\Omega_D
\end{equation}
\begin{equation}
s=K1+K2.
\end{equation}
 
\begin{equation}\nonumber
K1=-\frac{9a^3{H_0}^2 \Omega_{m0} H^2 \Omega_D^2\eta^5+k1}{3a\eta H[-9a^3{H_0}^2 \Omega_{m0}\eta^5 H^2\Omega_D+k11] }
\end{equation}
\begin{equation}\nonumber
k1=-2BHa^5\eta^{2\delta}\Omega_D^2(\delta-2)^2
\end{equation}
\begin{equation}\nonumber
k11=+B\eta^{2\delta}\left(-c^2 \eta+a^5 H(3a\eta H+2(\delta-2)\Omega_D)\right)
\end{equation}

\begin{equation}\nonumber
K2=-\frac{B\eta^{2\delta}\left(3a c^2 H \eta^2+k2\right)}{3a\eta H[-9a^3{H_0}^2 \Omega_{m0}\eta^5 H^2\Omega_D+k22] }
\end{equation}

\begin{equation}\nonumber
k2=(\delta-2)(c^2 \eta+2Ha^5(-5+2\delta+2aH\eta))\Omega_D
\end{equation}
\begin{equation}\nonumber
k22=B\eta^{2\delta}\left(-c^2 \eta+a^5 H(3a\eta H+2(\delta-2)\Omega_D)\right)
\end{equation}

\begin{figure}[htp]
\begin{center}
\includegraphics[width=6cm]{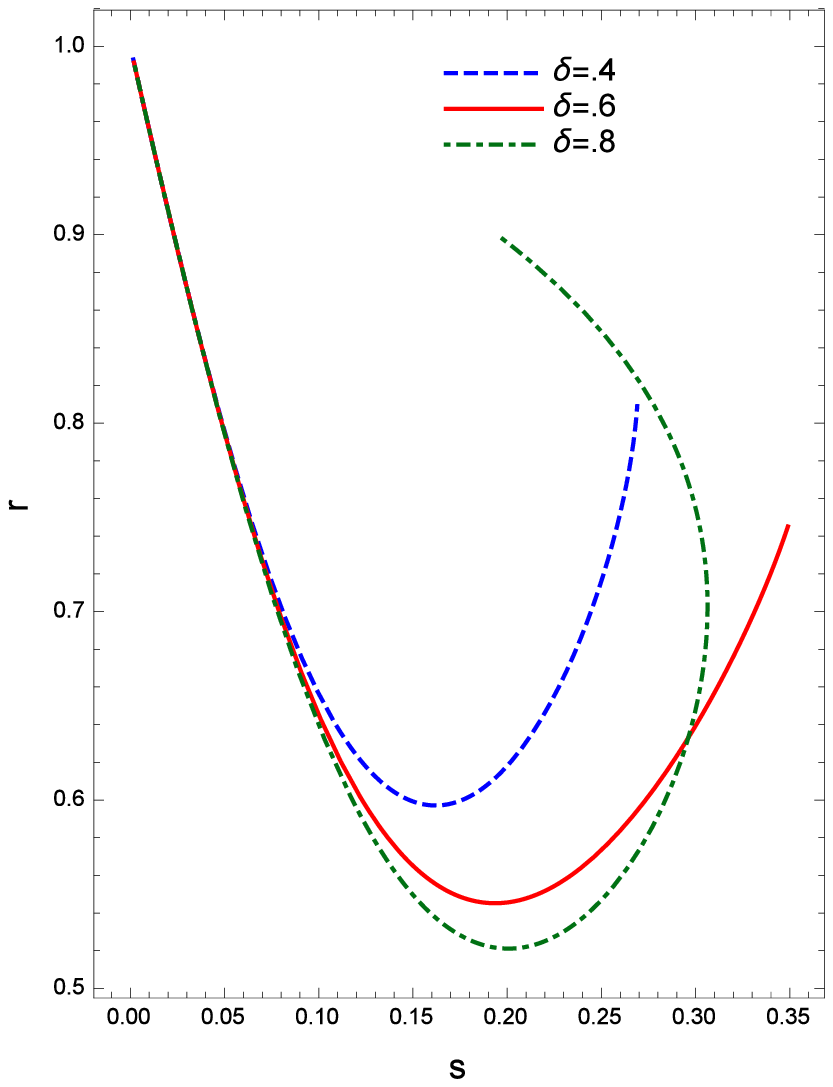}
\caption{The evolution of the statefinder parameter $r$ versus $s$ for
 non-interacting NTADE in CS modified gravity. Here, we have taken
$\Omega_D(z=0)=0.73$, $H(z=0)=74$, $\Omega_{m0}=.23$, $c^2=.25$ and $B=2.4$}\label{rs-z1}
\end{center}
\end{figure}

\begin{figure}[htp]
\begin{center}
\includegraphics[width=6cm]{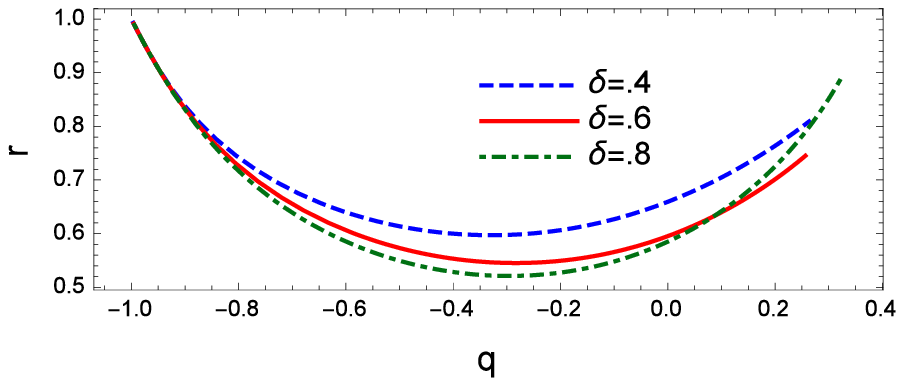}
\caption{The evolution of the statefinder parameter $r$ versus the
deceleration parameter $q$ for non-interacting NTADE in CS modified gravity. Here, we have taken
$\Omega_D(z=0)=0.73$, $H(z=0)=74$, $\Omega_{m0}=.23$, $c^2=.25$ and $B=2.4$}\label{rq-z1}
\end{center}
\end{figure}

\begin{figure}[htp]
\begin{center}
\includegraphics[width=8cm]{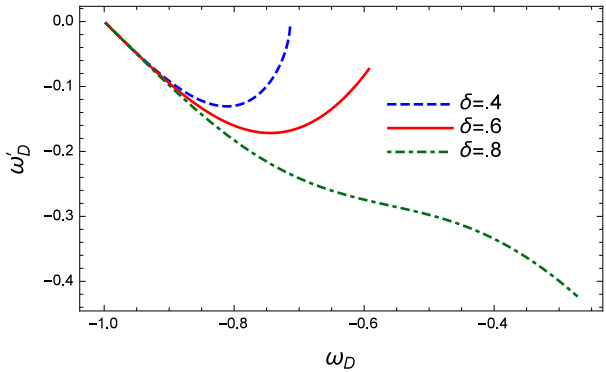}
\caption{The $\omega_D-{\omega}^{\prime}_{D}$ diagram for
non-interacting NTADE in CS modified gravity. Here, we have taken
$\Omega_D(z=0)=0.73$, $H(z=0)=74$, $\Omega_{m0}=.23$, $c^2=.25$ and $B=2.4$}\label{ww-z1}
\end{center}
\end{figure}

The evolution of $\Omega_D$ versus redshift parameter $z$, for non-interacting case is plotted in Fig.~\ref{Omega-z1}. From this figure we see that, at the late time, we have $\Omega_D\rightarrow 1$, which is consistent with observations. The equation of state (EoS) parameter given in Eq.(\ref{EoSna}) is also plotted in Fig.~\ref{w-z1} showing that, for different values of $\delta$, the EoS parameter displays quintessence-like nature of the universe. The behaviour of the deceleration parameter $q$ and the squared of sound speed $v_{s}^{2}$ are plotted in Figs.~\ref{q-z1}-\ref{v-z1}, which show, there is a deceleration expansion at the early time followed by an acceleration expansion and since $v_{s}^{2}<0$, the model is unstable, respectively. The cosmological planes for statefinder and  $\omega_D-{\omega}^{\prime}_{D}$, for non-interacting case are plotted in Fig.~\ref{rs-z1}-\ref{ww-z1}. In Fig.~\ref{ww-z1}, we see for $\omega_D<0$, the evolution parameter shows ${\omega}^{\prime}_{D}<0$, which represents the freezing region of evolving universe.

\subsection{Interacting case}
Recent observations indicate that the evolution of DM and DE is not independent, rather, it is a response to the coincidence problen \cite{maz5}. Thus, we consider the interaction between the two fluids. At here, by considering the form of interaction as $Q=3 H b^{2}\rho_{m}$, we study the behavior of the cosmological parameters of the model. Like the previous section, by matter of calculations we can find expressions for the cosmological parameters, $\Omega_D$, $\omega_D$, $q$, and $v_{s}^{2}$ as
\begin{figure}[htp]
\begin{center}
\includegraphics[width=8cm]{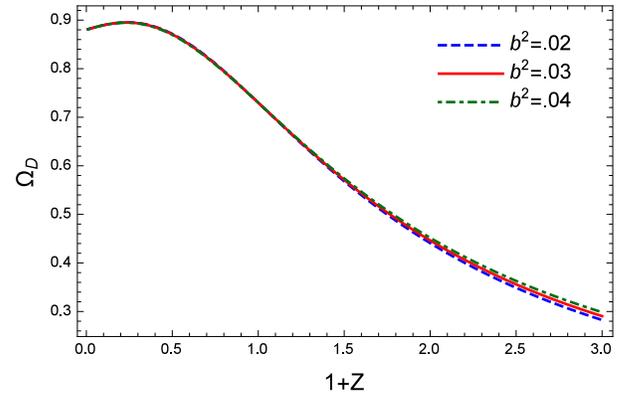}
\caption{The evolution of $\Omega_D$ versus redshift parameter $z$ for
 interacting NTADE in CS modified gravity. Here, we have taken
$\Omega_D(z=0)=0.73$, $H(z=0)=74$, $\Omega_0=.23$, $c^2=.25$, $B=2.4$ and $\delta=.6$
}\label{Omega-z2}
\end{center}
\end{figure}

\begin{figure}[htp]
\begin{center}
\includegraphics[width=8cm]{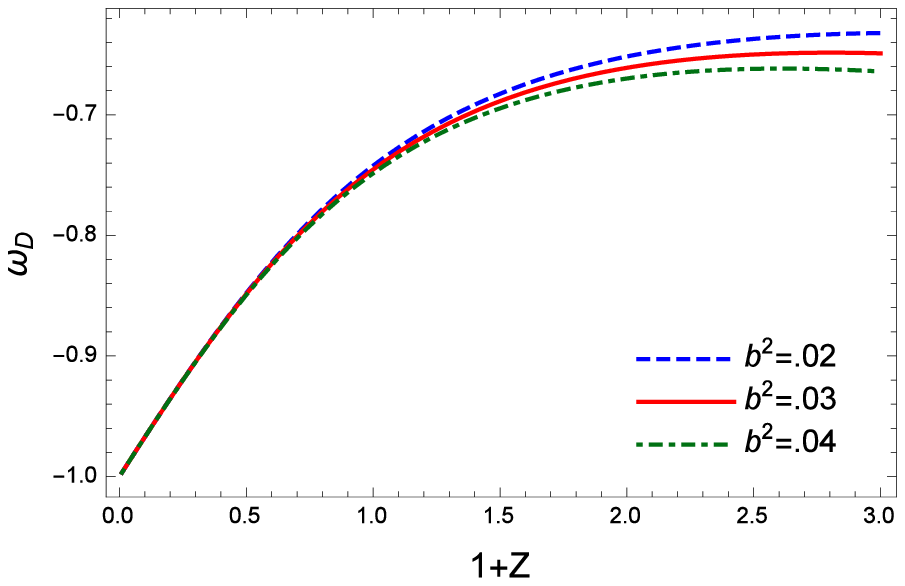}
\caption{The evolution of $\omega_D$ versus redshift parameter $z$ for
 interacting NTADE in CS modified gravity. Here, we have taken
$\Omega_D(z=0)=0.73$, $H(z=0)=74$, $\Omega_{m0}=.23$, $c^2=.25$, $B=2.4$ and $\delta=.6$}\label{w-z2}
\end{center}
\end{figure}

\begin{figure}[htp]
\begin{center}
\includegraphics[width=8cm]{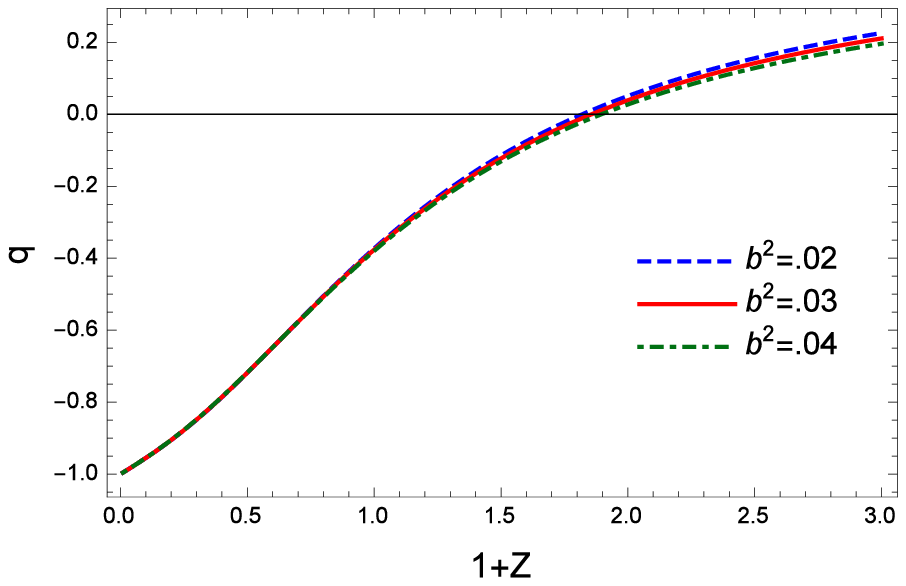}
\caption{The evolution of the deceleration parameter $q$ versus redshift parameter $z$ for
 interacting NTADE in CS modified gravity. Here, we have taken
$\Omega_D(z=0)=0.73$, $H(z=0)=74$, $\Omega_{m0}=.23$, $c^2=.25$, $B=2.4$ and $\delta=.6$}\label{q-z2}
\end{center}
\end{figure}

\begin{figure}[htp]
\begin{center}
\includegraphics[width=8cm]{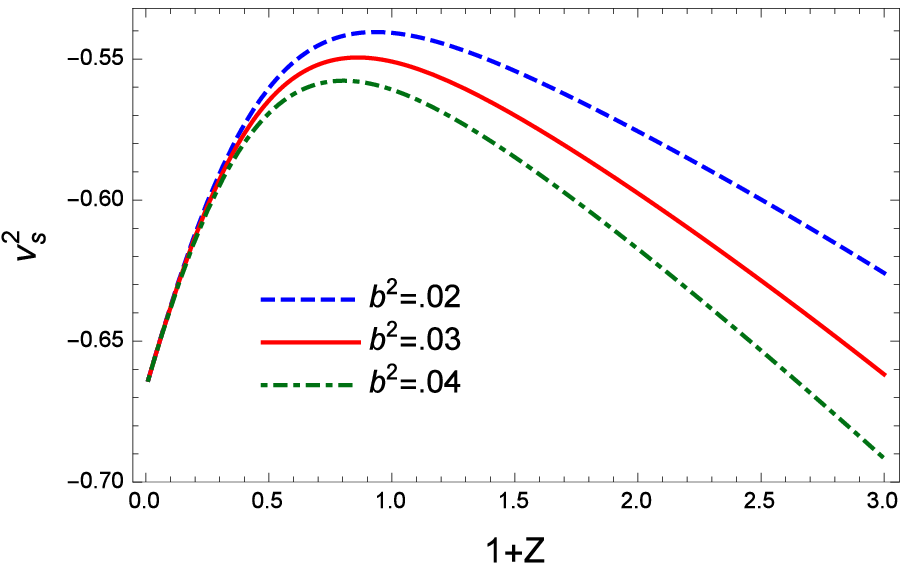}
\caption{The evolution of  the squared of sound speed $v_s^2 $ versus redshift parameter $z$ for
 interacting NTADE in CS modified gravity. Here, we have taken
$\Omega_D(z=0)=0.73$, $H(z=0)=74$, $\Omega_{m0}=.23$, $c^2=.25$, $B=2.4$ and $\delta=.6$}\label{v-z2}
\end{center}
\end{figure}

\begin{figure}[htp]
\begin{center}
\includegraphics[width=6cm]{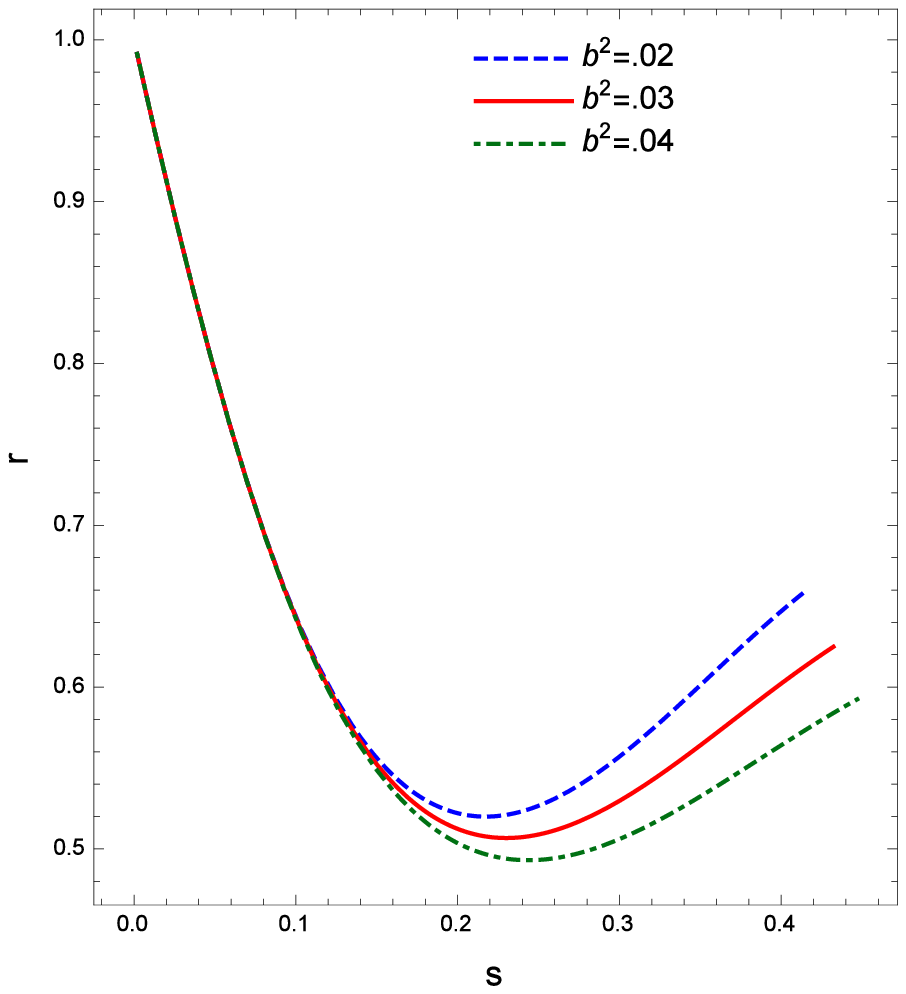}
\caption{The evolution of the statefinder parameter $r$ versus $s$ for
 interacting NTADE in CS modified gravity. Here, we have taken
$\Omega_D(z=0)=0.73$, $H(z=0)=74$, $\Omega_{m0}=.23$, $c^2=.25$, $B=2.4$ and $\delta=.6$}\label{rs-z2}
\end{center}
\end{figure}

\begin{figure}[htp]
\begin{center}
\includegraphics[width=6cm]{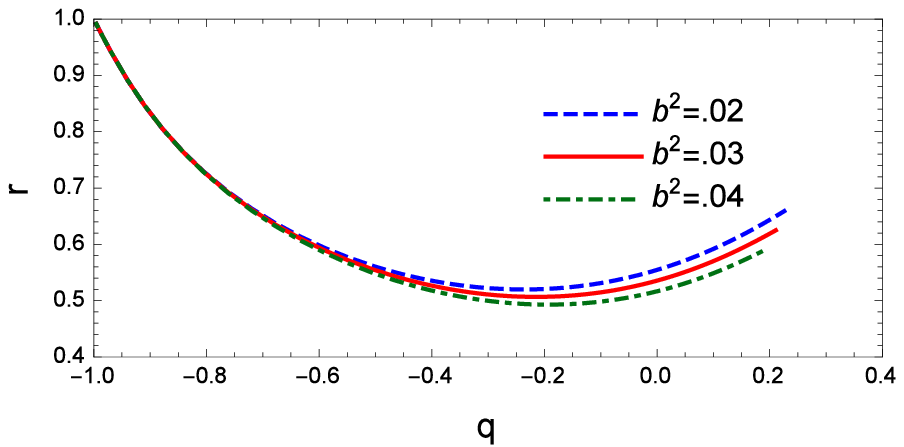}
\caption{The evolution of the statefinder parameter $r$ versus the
deceleration parameter $q$ for interacting NTADE in CS modified gravity. Here, we have taken
$\Omega_D(z=0)=0.73$, $H(z=0)=74$, $\Omega_{m0}=.23$, $c^2=.25$, $B=2.4$ and $\delta=.6$}\label{rq-z2}
\end{center}
\end{figure}

\begin{figure}[htp]
\begin{center}
\includegraphics[width=8cm]{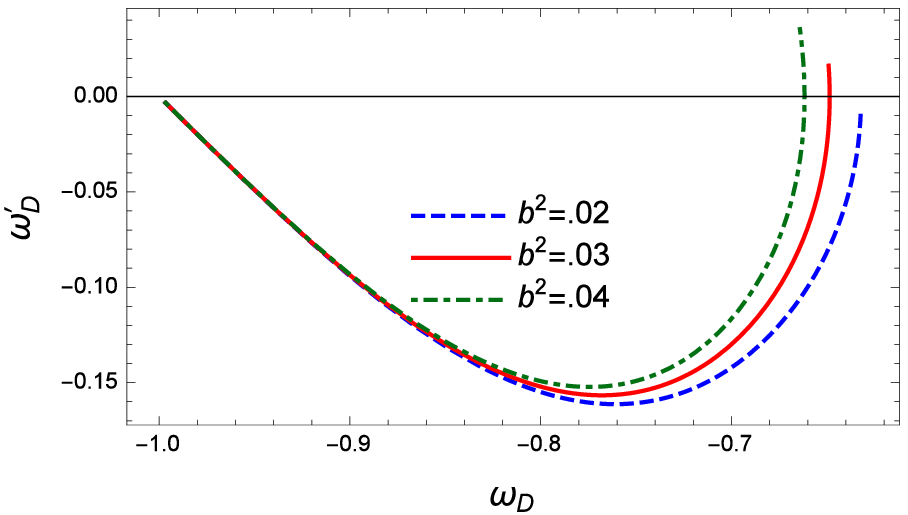}
\caption{The $\omega_D-{\omega}^{\prime}_{D}$ diagram for
 interacting NTADE in CS modified gravity. Here, we have taken
$\Omega_D(z=0)=0.73$, $H(z=0)=74$, $\Omega_{m0}=.23$, $c^2=.25$, $B=2.4$ and $\delta=.6$}\label{ww-z2}
\end{center}
\end{figure}

\begin{figure}[htp]
\begin{center}
\includegraphics[width=8cm]{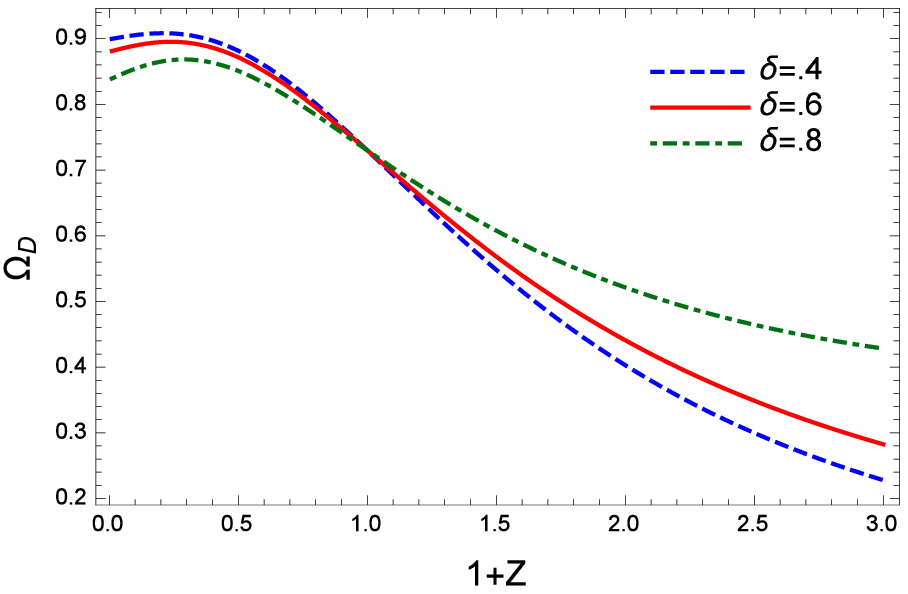}
\caption{The evolution of $\Omega_D$ versus redshift parameter $z$ for
 interacting NTADE in CS modified gravity. Here, we have taken
$\Omega_D(z=0)=0.73$, $H(z=0)=74$, $\Omega_{m0}=.23$, $c^2=.25$, $B=2.4$ and $b^2=.02$
}\label{Omega-z3}
\end{center}
\end{figure}

\begin{figure}[htp]
\begin{center}
\includegraphics[width=8cm]{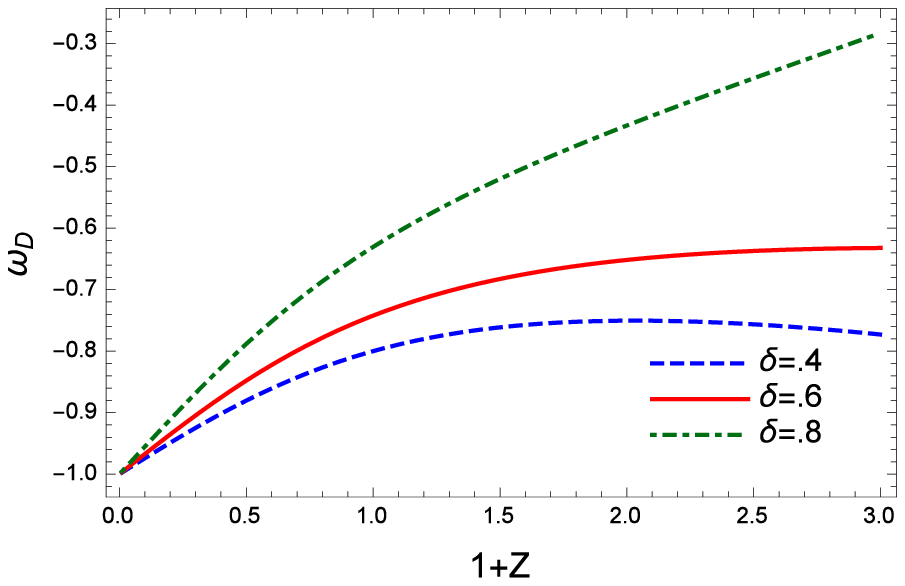}
\caption{The evolution of $\omega_D$ versus redshift parameter $z$ for
 interacting NTADE in CS modified gravity. Here, we have taken
$\Omega_D(z=0)=0.73$, $H(z=0)=74$, $\Omega_{m0}=.23$, $c^2=.25$, $B=2.4$ and $b^2=.02$}\label{w-z3}
\end{center}
\end{figure}

\begin{figure}[htp]
\begin{center}
\includegraphics[width=8cm]{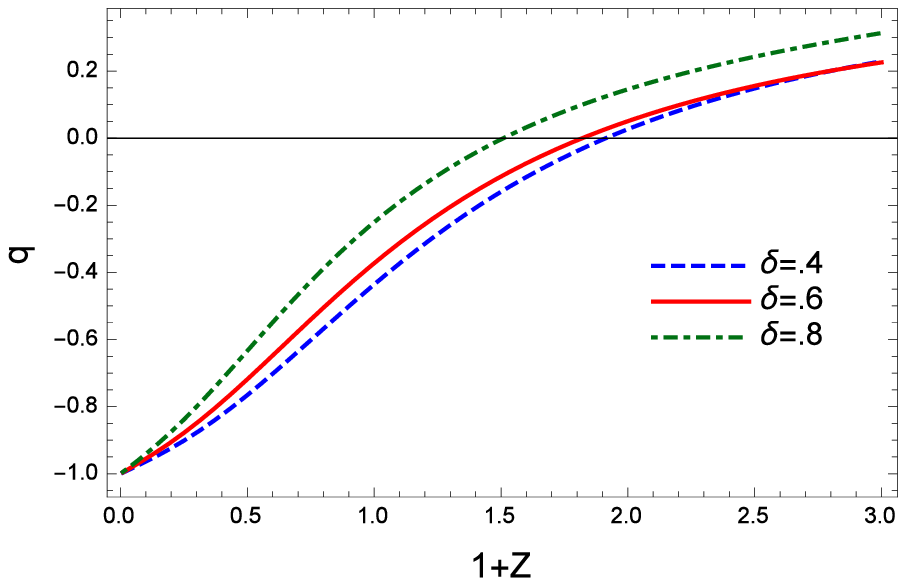}
\caption{The evolution of the deceleration parameter $q$ versus redshift parameter $z$ for
 interacting NTADE in CS modified gravity. Here, we have taken
$\Omega_D(z=0)=0.73$, $H(z=0)=74$, $\Omega_{m0}=.23$, $c^2=.25$, $B=2.4$ and $b^2=.02$}\label{q-z3}
\end{center}
\end{figure}

\begin{figure}[htp]
\begin{center}
\includegraphics[width=8cm]{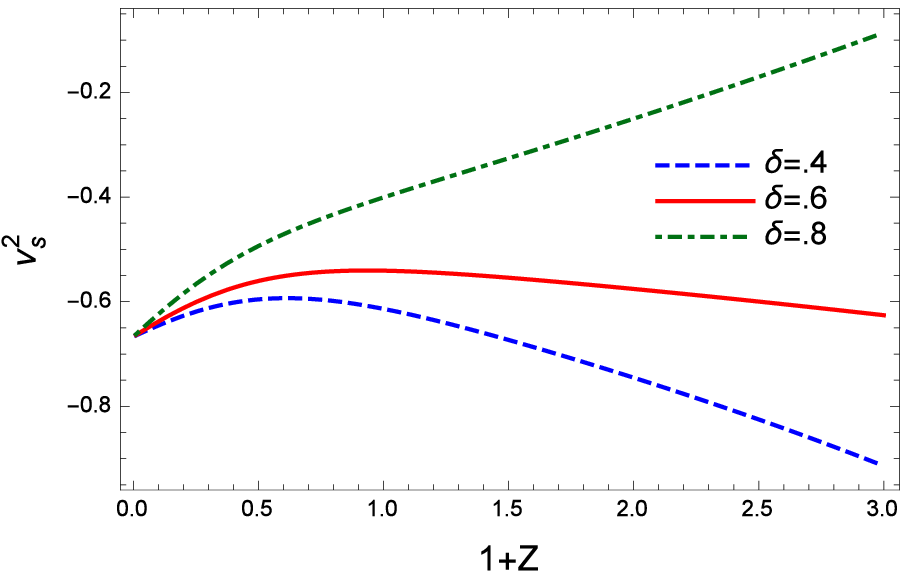}
\caption{The evolution of  the squared of sound speed $v_s^2 $ versus redshift parameter $z$ for
 interacting NTADE in CS modified gravity. Here, we have taken
$\Omega_D(z=0)=0.73$, $H(z=0)=74$, $\Omega_{m0}=.23$, $c^2=.25$, $B=2.4$ and $b^2=.02$}\label{v-z3}
\end{center}
\end{figure}

\begin{figure}[htp]
\begin{center}
\includegraphics[width=6cm]{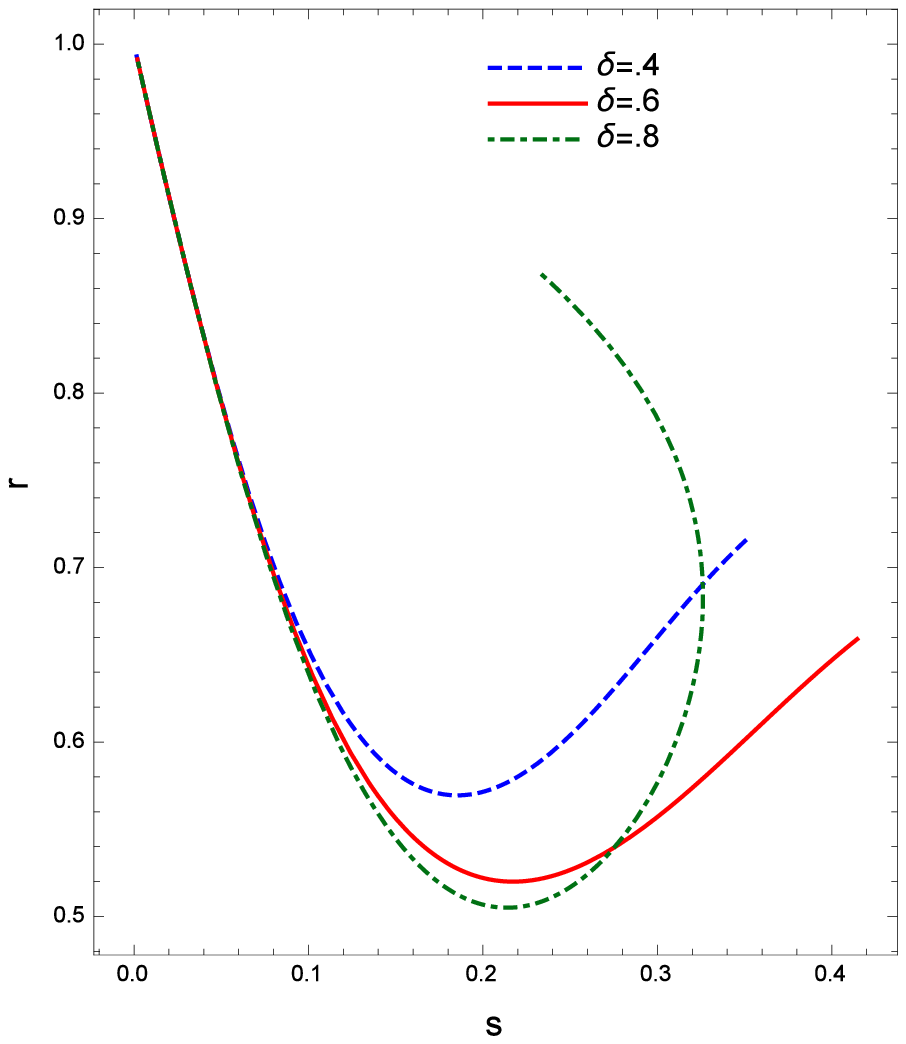}
\caption{The evolution of the statefinder parameter $r$ versus $s$ for
 interacting NTADE in CS modified gravity. Here, we have taken
$\Omega_D(z=0)=0.73$, $H(z=0)=74$, $\Omega_{m0}=.23$, $c^2=.25$, $B=2.4$ and $b^2=.02$}\label{rs-z3}
\end{center}
\end{figure}

\begin{figure}[htp]
\begin{center}
\includegraphics[width=6cm]{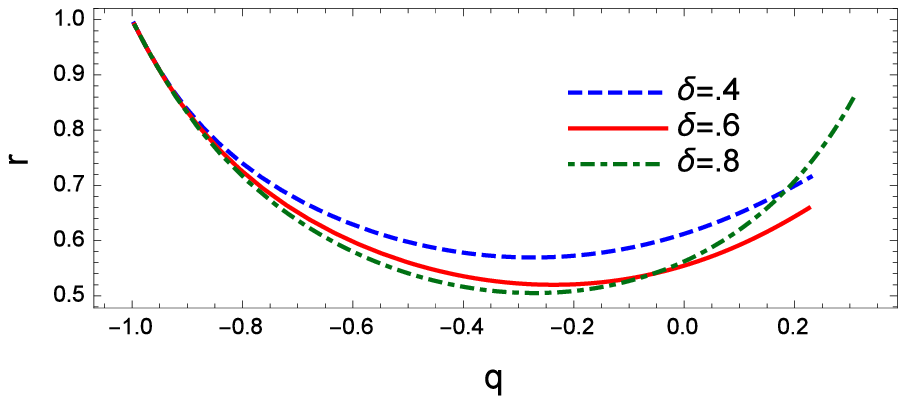}
\caption{The evolution of the statefinder parameter $r$ versus the
deceleration parameter $q$ for interacting NTADE in CS modified gravity. Here, we have taken
$\Omega_D(z=0)=0.73$, $H(z=0)=74$, $\Omega_{m0}=.23$, $c^2=.25$, $B=2.4$ and $b^2=.02$}\label{rq-z3}
\end{center}
\end{figure}

\begin{figure}[htp]
\begin{center}
\includegraphics[width=8cm]{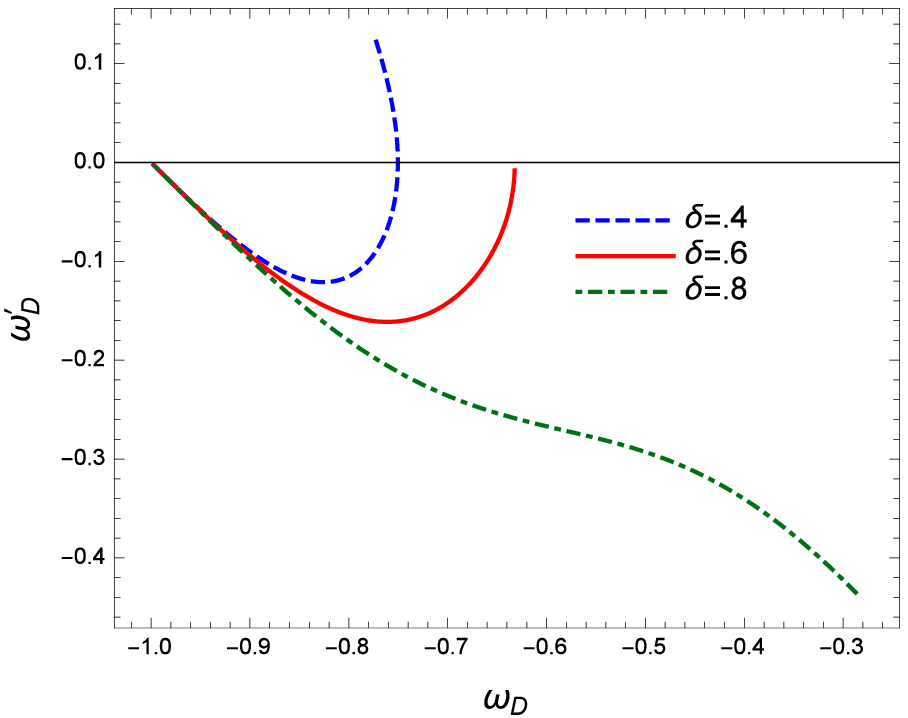}
\caption{The $\omega_D-{\omega}^{\prime}_{D}$ diagram for
 interacting NTADE in CS modified gravity. Here, we have taken
$\Omega_D(z=0)=0.73$, $H(z=0)=74$, $\Omega_{m0}=.23$, $c^2=.25$, $B=2.4$ and $b^2=.02$}\label{ww-z3}
\end{center}
\end{figure}


\begin{eqnarray}\label{EoSna1}
\omega_D=-1-b^2 u-\frac{2\delta-4}{3a\eta H},
\end{eqnarray}

\begin{equation}\nonumber
{\omega}^{\prime}_{D}=\frac{-3 a^{3+3b^2}H_0^2\Omega_{m0} \eta^{3-2\delta}(3a(b^2-1)b^2 \eta H-(\delta-2)w1))}{a^7 B H}
\end{equation}
\begin{equation}\label{primeEoS1}
+\frac{(\delta-2)(-c^2 \eta+2a^5 H(1+a\eta H+(\delta-2)\Omega_D))}{3a^7 \eta^2 H^3}
\end{equation}
\begin{equation}\nonumber
w1=(2b^2+(b^2-1)\Omega_D
\end{equation}
\begin{equation}\label{qnage1}
q=-1-\Omega_D(\frac{3u(b^2-1)}{2}+\frac{\delta-2}{a\eta H})+\frac{c^2 a^{-6}}{2H^2},
\end{equation}
\begin{equation}\label{nageOmega1}
 {\Omega}^{\prime}_{D}=2\Omega_D(\frac{\delta-2}{a\eta H}+1+q).
\end{equation}
\begin{equation}\nonumber
v_{s}^{2}=\frac{-3a^{3+3b^2}(b^2-1) {H_0}^2 \Omega_{m0} \eta^{4-2\delta} (3a b^2 \eta H -(\delta-2)\Omega_D)}{2a^6 B(\delta-2)}
\end{equation}
\begin{equation}\label{vs2}
+\frac{-c^2 \eta+2a^5 H(5-2\delta-2a\eta H+(\delta-2)\Omega_D)}{6 a^6 \eta H^2},
\end{equation}
where $u=\frac{\rho_m}{\rho_D}$ is the ratio of the energy densities. 
Since the expression of $r$ and $s$ are too long, we do not present them here.


The cosmological parameters, like, $\Omega_D$, $\omega_D$, $q$ and $v_{s}^{2}$ for interacting case with different values of $b^2$ are plotted in Figs.~\ref{Omega-z2}-\ref{v-z2}. It is obvious from these figures that  the model shows a quintessence-like behaviour for the EoS parameter. Additionally, the universe experiences a transition from a deceleration phase to an acceleration one despite the instability. Also, for different values of the coupling constant $b^2$, the cosmological planes are plotted in Figs.~\ref{rs-z2}-\ref{ww-z2}. In Fig.~\ref{rs-z2}, we see trajectories $r$ and $s$ end at $\Lambda CDM$ $(r=1, s=0)$ for different values of $b^2$ at low redshift of NTADE in CS modified gravity as well as in Fig.~\ref{rq-z2}, we observe for $(r, q)$ evolutionary plane, the evolutionary trajectories stated from matter dominated universe in the past and approach the point $(r=1, q=-1)$ in the future. The evolutionary trajectories of $\omega_D-{\omega}^{\prime}_{D}$ plane are shown in Fig.~\ref{ww-z2}, with different values of $b^2$. We see for NTADE in CS modified gravity, $\omega_D$ behaves like quintessence type dark energy with $\omega_D > -1$, as well as the value of ${\omega}^{\prime}_{D}$, at first decreases to minimum then increases to zero in the future.

The plots of $\Omega_D$, $\omega_D$,$q$ , $v_{s}^{2}$ and $\omega_D-{\omega}^{\prime}_{D}$ for a fixed $b^2$ and different values of $\delta$ are shown in Figs.~\ref{Omega-z3}-\ref{ww-z3}. It can be observed from Fig. \ref{w-z3} that the EoS parameter exhibits the quintessence-like behaviour for the different values of $\delta$ as well as by considering different values of $\delta$, Fig. \ref{v-z3} shows this models is unstable. 
 The evolutionary trajectories for $(r-s)$ and $(r-q)$ planes for NTADE model in CS modified gravity have plotted in Figs. \ref{rs-z3} and \ref{rq-z3} respectively. We see from Fig.  \ref{rs-z3} that, the evolutionary trajectories $r$ and $s$ end at $\Lambda CDM$ $(r=1, s=0)$ in the future for different values of $\delta$ as well as in Fig. \ref{rq-z3} , we see for NTADE in CS modified gravity, the evolutionary trajectories started from matter dominated universe in the past and approach the point $(r=1, q=-1)$ in the future. 
The $\omega_D-{\omega_D}^{\prime}$ plane for NTADE model in CS modified gravity has plotted in Fig.  \ref{ww-z3} which represents the freezing region.

\section{Closing remarks}
In this work, we studied the behaviour of cosmological parameters and cosmlogical planes for NTADE in the framework of the Chern-Simons modified gravity model in spatially flat universe for non-interacting and interacting cases. The graphical behaviour of squared of sound speed and the EoS parameter are displayed against redshift parameter in this paper. We see, since the value of $v_{s}^{2}$ is negative, the model is not stable and the equation of state parameter presents quintessence- like nature of the universe. The $\omega_D-{\omega}^{\prime}_{D}$
plane for the present scenario is displayed in this paper, which exhibit the freezing region for most of cases.


\end{document}